\documentclass[9pt,twocolumn]{article}

\usepackage[margin=0.5 in]{geometry}
\usepackage{amsmath}
\usepackage{amsfonts,amssymb}
\usepackage{graphicx}
\usepackage[T1]{fontenc}
\usepackage[latin1]{inputenc}
\usepackage{caption}
\usepackage{cite}

\title{Study of reflectors for illumination via conformal maps}

\author{Luis A. Alem\'an-Casta\~neda$^{1,2,*}$ and Miguel A. Alonso$^{1,2,3,**}$}
\date{
	\small \textit{
	$^1$The Institute of Optics and Center for Freeform Optics, University of Rochester, Rochester NY 14627, USA.\\%
	$^2$Aix-Marseille Univ., Institut Fresnel, UMR 7249, 13397 Marseille Cedex 20, France.\\
	$^3$CNRS, \'Ecole Centrale Marseille,  13013 Marseille, France.\\
	$^*$lalemanc@ur.rochester.edu, $^{**}$miguel.alonso@fresnel.fr\\
	[2ex]%
	\today}
}

\begin{document}

\twocolumn[
\begin{@twocolumnfalse}
	\maketitle
	\begin{abstract}
		We present an approach for the study and design of reflectors with rotational or translational symmetry that redirect light from a point source into any desired radiant intensity distribution. This method is based on a simple conformal map that transforms the reflectors' shape into a curve that describes light's direction after reflection. 
		Both segmented and continuous reflectors are discussed, illustrating how certain reflector characteristics become apparent under this transformation. 
		This method can also be used to study extended sources via translations. 
	\end{abstract}
\end{@twocolumnfalse}
]

Early work on illumination optics consisted on reflector design through source-target maps, i.e. the relationship between the ray angles from the source to the reflector and the target direction, and through parameters describing the system, such as the acceptance angle \cite{winston:selectedpapers}.
State-of-the-art techniques have shifted the focus to assessing/obtaining a given irradiance or radiant intensity distributions, and great progress has been achieved in both numerical and analytical implementations \cite{Minano:92,Ries:02,Oliker:03,Canavesi:13,Wester:14,Hirst:18}. 
Comprehensive reviews can be found in \cite{winston:selectedpapers,Winston:18,Wu:18}. The general 3D problem is governed by a nonlinear second-order 
Monge-Ampere equation, which usually must be solved numerically.  
When there is either rotational or translational symmetry, however, the system reduces to a 2D problem, and analytic methods can then be used, e.g, the so-called 
``edge-ray'' \cite{Gordon:93,Davies:94,Ries:94,Rabl:94,Ong:96}, and``flow-line'' \cite{Jiang:16,Jiang_Symmetric:16,Winston:18} methods. The method proposed here falls in this category. 

Point-source (or zero-\'etendue) algorithms are of great importance despite their limited validity. Point sources play a role in many general schemes for designing reflectors with extended sources, such as the edge-ray method. 
The supporting paraboloids \cite{Kochengin:1997,Canavesi_Direct:12} and linear programming \cite{Canavesi_Linear:12,Canavesi:13} methods have been successful in generating desired intensity distributions with good numerical performance, even when optimizing for extended sources. In these two methods, the reflector is subdivided into patches of paraboloids so that, based on the properties of conics, a far-field intensity distribution is tailored. Although these properties have been used extensively in non-imaging optics, the connection between their shape and the direction of light after reflection has been restricted mainly to piecewise parabolic reflectors. Our aim is to provide a simple conceptual connection valid also in the continuous, smooth case, hence providing a simple, visual tool for the assessment of any reflector with rotational or translational symmetry. 

The usefulness and elegance of the transformation proposed here is illustrated through a series of examples. The first is a simple piecewise parabolic reflector to help introduce the conformal map and the relation between the reflector and its transformed curve. The second example shows how, by employing this  map, a piecewise parabolic reflector can be easily designed. The next two examples --also piecewise parabolic-- exhibit how reflectors can be studied via this transformation, and, specifically, how their geometry is revealed --whether light converges, diverges or is collimated. The last two examples correspond to continuous, smooth reflectors used as collimators for point-like sources and as collectors for extended sources. 

The treatment that follows is two-dimensional. Consider a parabolic reflector of focal length $f$ whose focus is at the origin and  whose axis is at an angle $\beta$ with respect the $x$-axis. Light from a point source at the origin is redirected parallel to the parabola's axis. The goal is to find a map that transforms such parabolas into straight lines whose directions are related to that of the redirected light. 
It is useful to write in polar coordinates the general expressions of both the parabola and a straight line with impact parameter (distance to the origin) $\bar{d}$ and angle $\bar{\beta}$:
\begin{eqnarray}
\label{eq:polar_exp}
\text{parabola: }r(\theta)=f\csc^2\!\left(\frac{\theta-\beta}{2}\right)\\ {\text{line: }}\bar{r}(\bar{\theta})=\bar{d}\csc\!\left(\bar{\theta}-\bar{\beta}\right),
\end{eqnarray}
where an overbar is used to label quantities that refer to the straight line, to distinguish them from those of the parabola. It follows that the transformations
\begin{equation}\label{eq:square_maps}
\begin{aligned}
\theta=2\bar{\theta}, & & & r=\bar{r}^2,
\end{aligned}
\end{equation}
provide the desired map that converts a line with parameters $(\sqrt{f},\beta/2)$ into a parabola with parameters $(f,\beta)$. In other words, if we express the reflector's 2D coordinates $(x,y)$ as a complex number $z=x+iy$, the conformal map $\sqrt{z}=\bar{z}=\bar{x}+i\bar{y}$ exactly transforms parabolas with focus at the origin into straight lines. This map has been used in other physical contexts, e.g. to connect Keplerian motion  whose orbits are ellipses or hyperbolas centered at one of the foci, with the 
motion in a symmetric 2D quadratic (attractive or repulsive) potential whose orbits are ellipses or hyperbolas centered at the potential's center  \cite{Mittag:92}.  
The case considered here of a parabola being mapped onto a straight line corresponds to the limiting case where the quadratic potential vanishes (free motion) and the Keplerian path is an escape/capture orbit. 
This map has also been used to describe the caustics of structured Gaussian beams \cite{Alonso:17}.

\begin{figure}[!ht]
\centering
\includegraphics[width=0.94\linewidth]{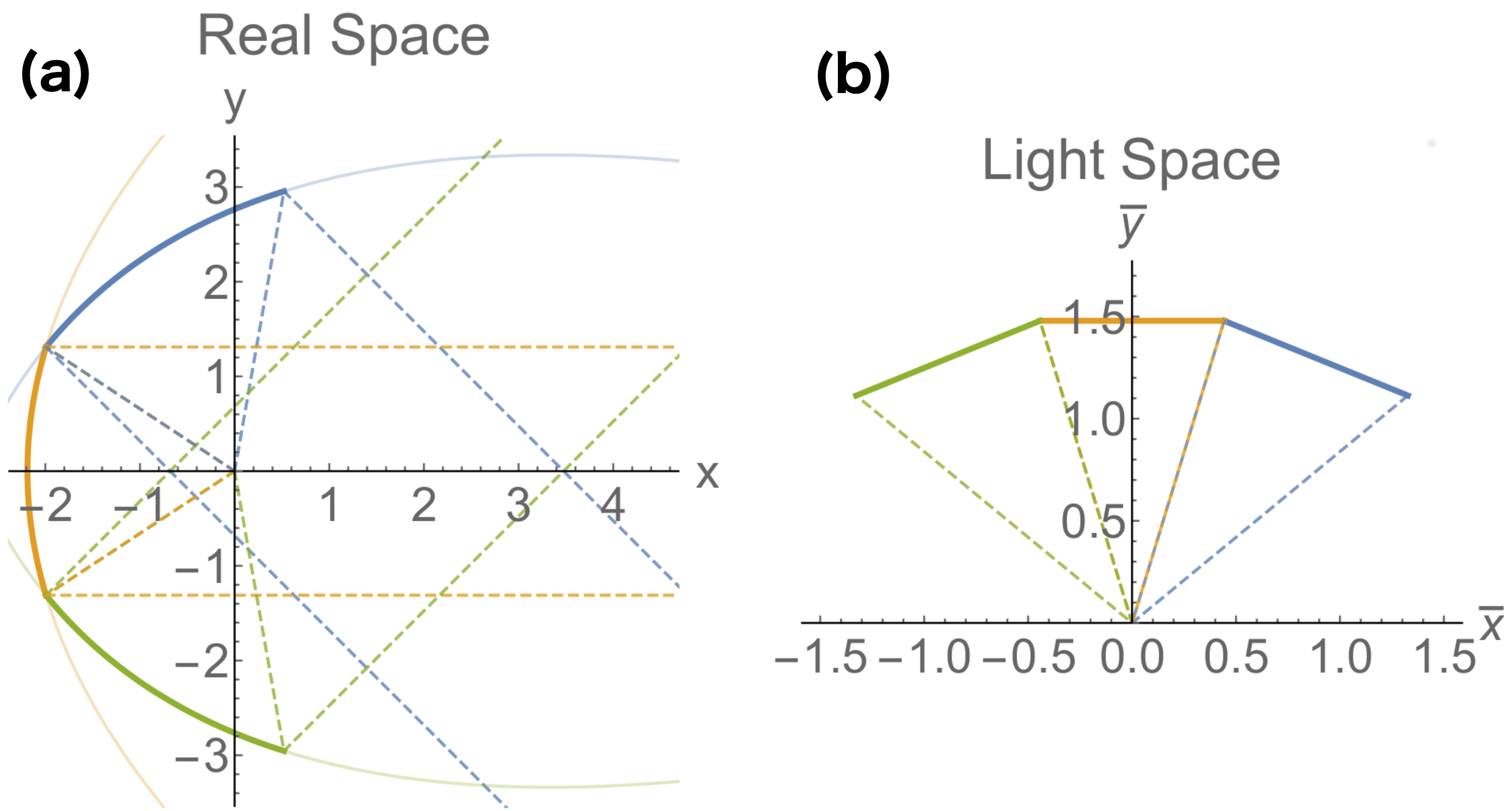}
\caption{\footnotesize (a) Segmented reflector consisting of three parabolic segments, each depicted in a different color. 
The limiting rays are shown as dashed lines. Each segment redirects the same amount of power from the source. 
(b) The corresponding light curve is composed of three straight line segments at angles from the $\bar{x}$-axis equal to half the reflected light's angles from the $x$-axis. 
Although the line segments' lengths are different, they subtend equal angles from the origin, which are half the angle subtended by the parabolic segments.}
\label{fig:first_example}
\end{figure}

We can then define an abstract \textit{light space} with coordinates $(\bar{x},\bar{y})$ in which a \textit{light curve} exists, in contrast with the \textit{real space} with coordinates $(x,y)$ where the reflector exists. These spaces are connected by the map just described. This is illustrated in Fig.~\ref{fig:first_example} for 
a continuous reflector composed of contiguous confocal parabolic segments
--as in the supporting paraboloids method. The resulting light curve
is simply a chain of straight line segments, where the angle that each segment makes with the $\bar{x}$-axis is just half the angle of the reflected light with respect to the $x$-axis, and the angle subtended from the origin by each segment (not its length) is proportional to the corresponding redirected power. Because the map $\bar{z}=\sqrt{z}$ is not single valued, a branch cut must be chosen, and it is convenient for continuity of the light curve to choose it as the average target direction. However, the choice of branch changes neither the direction nor the subtended angles.

\begin{figure}[htpb]
\centering
\includegraphics[width=0.92\linewidth]{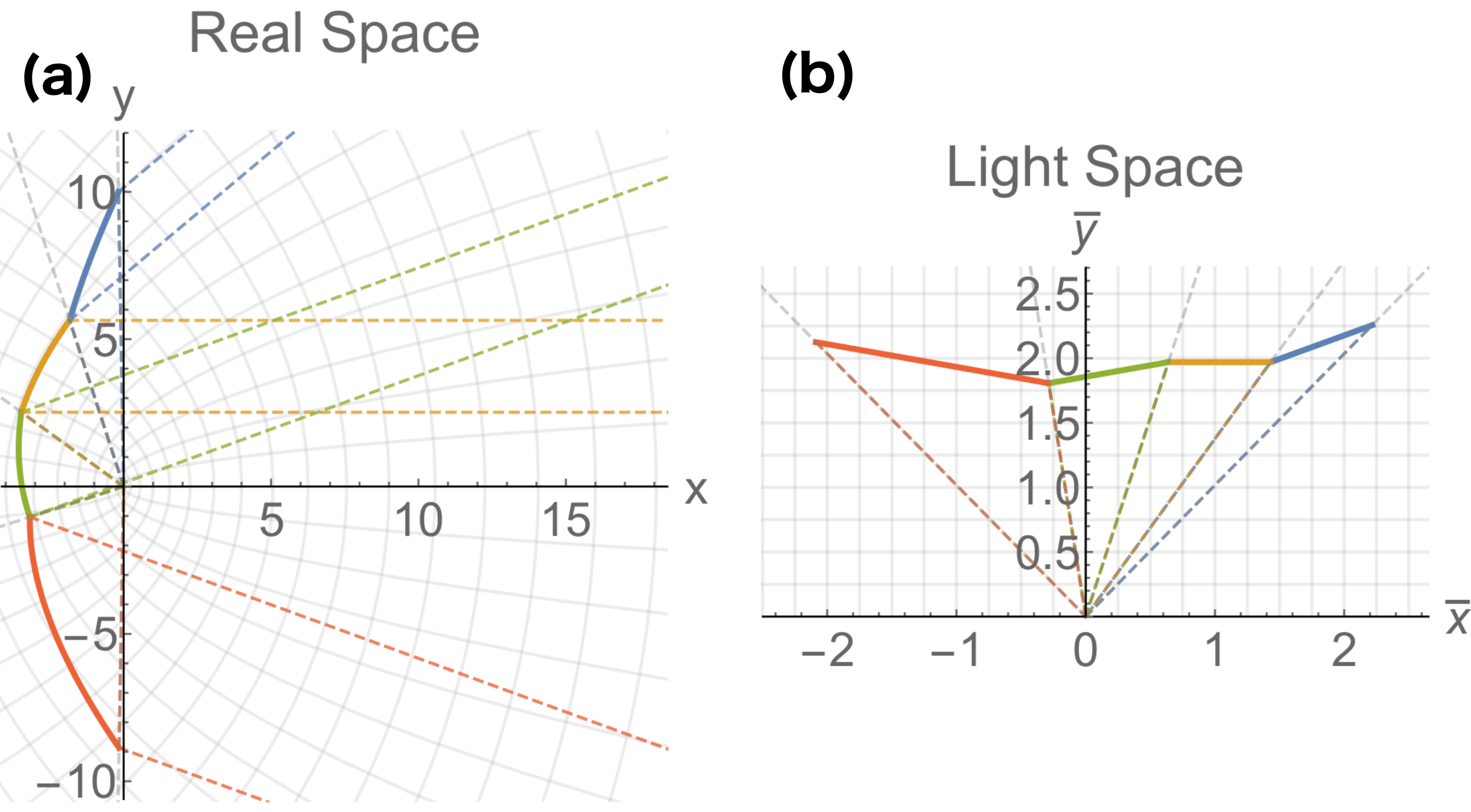}
\caption{\footnotesize (a) The reflector is obtained after applying the $z=\bar{z}^2$ map. (b) Light curve construction given a source emitting uniformly and a source range $\alpha\in[90^\circ,270^\circ]$, i.e. from $45^\circ$ to $135^\circ$ in the light space. We want to redirect  $10\%,~20\%,~30\%,$ and $40\%$ of light towards $\beta=40^\circ,~0^\circ,~20^\circ$, and $-20^\circ$ respectively, so we draw the corresponding straight-line segments oriented at half those target angles. (See {\bf Visualization 1}.)}
\label{fig:design}
\end{figure}

When designing a reflector in 2D, one must choose the source-target mapping, i.e. set the ray angles from the source to the reflector, $\alpha$, and the target direction, $\beta(\alpha)$. Both angles are measured from the optic axis, chosen here as the $x$-axis. Note that a source-target mapping does not determine a unique reflector but a family of solutions. 
The inverse map $z=\bar{z}^2$ can be used to design reflectors from their respective light curves without explicitly tracing any rays. For a piecewise parabolic reflector the procedure is the following: (1) given a light source, choose the source range; (2) in light space this defines a range subtending just half the angle, which must be subdivided into sections depending on the fraction of energy desired for each; (3) draw a straight-line segment oriented in each section at half the desired target angle --this constitutes the light curve; (4) obtain the reflector by applying the map $z=\bar{z}^2$. Note that connecting at their ends the light curve segments leads to a continuous reflector, while disconnected segments lead to  Fresnel-type reflectors. In Fig. \ref{fig:design} we construct a reflector following the previous procedure.

\begin{figure}[htpb]
\centering
\includegraphics[width=0.93\linewidth]{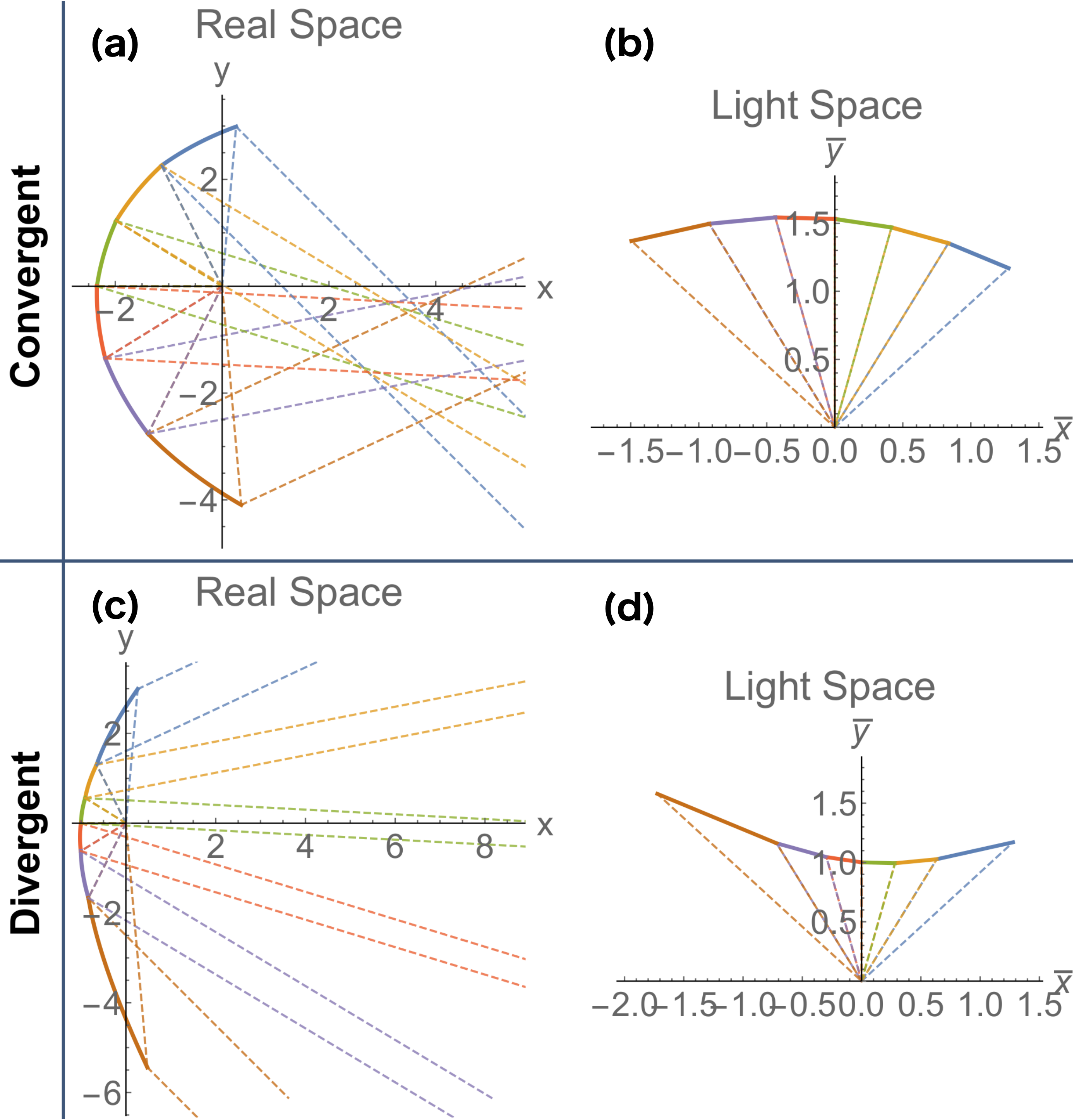}
\caption{\footnotesize Two reflectors composed of six parabolic segments (a),(c) and their light curves (b),(d). The reflectors have equal source and target ranges, but have convergent (a) and divergent (c) geometry. 
The light curves are concave/convex  towards the origin for convergent/divergent reflectors.}
\label{fig:segmented}
\end{figure}

In particular, there are two main geometries for the design of reflectors: convergent (where rays cross after reflection), and divergent (where they do not cross). Depending on the problem at hand, each geometry has advantages and disadvantages. 
The proposed conformal map reveals these geometries directly, as illustrated for the two piecewise-parabolic continuous reflectors shown in 
Fig.~\ref{fig:segmented} (with equal target and source ranges).
A concave/convex light curve towards the origin implies a convergent/divergent reflector. Note that light curves consisting of straight segments in the same directions, each subtending the same angles, yield after the  $z=\bar{z}^2$ map different reflectors  with the same source and target ranges, not necessarily purely convergent or divergent, as in Fig. \ref{fig:design}.

Let us now consider discontinuous reflectors such as piecewise-parabolic Fresnel-type reflectors, commonly used for collimation 
when thin reflectors are preferred due to packaging or construction limitations. 

\begin{figure}[htbp]
\centering
\includegraphics[width=0.98\linewidth]{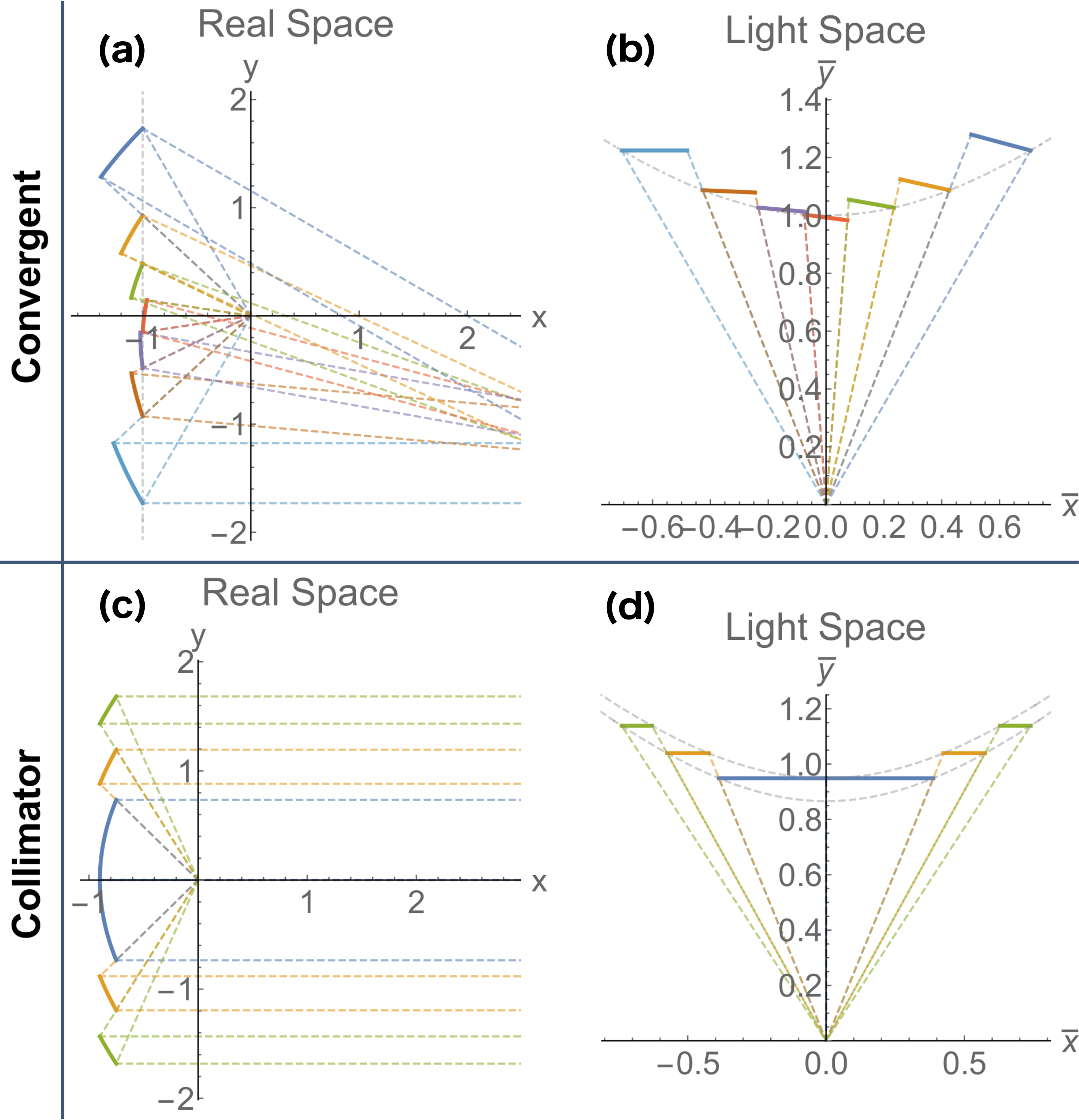}
\caption{\footnotesize Segmented Fresnel-type reflectors: (a) convergent reflector with seven segments and (b) its light curve. (c) collimator, (i.e. neither divergent or convergent) with five segments and (d) its light curve. The baselines for the reflector segments map onto limiting rectangular hyperbolas for the light curves.}
\label{fig:fresnel}
\end{figure}

Like the reflector, the light curve is also discontinuous, but since typically all reflector segments have a common substrate, the light curve segments also touch the mapped substrate. For example, if we consider that all segments start at a line oriented at $\theta_c$ and with impact factor $d$, then the light curves all depart from a rectangular hyperbola (with perpendicular asymptotes) oriented at $\theta_c/2-\pi/4$ and semi-axis $\sqrt{d}$. 
For the collimator, which lies between two lines, its corresponding segments in light space lie between two rectangular hyperbolas. 
Figure~\ref{fig:fresnel} shows two reflectors and their respective light curves. For a collimator, all light curve segments are parallel but disconnected. 
Note that while the light line segments  follow a hyperbola (a convex curve), the geometry of the reflector is recovered from the angular change of the light curve: convergent if co-rotating with respect to the origin, divergent if counter-rotating, and collimating if constant.

This approach is valid not only for segmented piecewise-parabolic reflectors. More general reflectors can be regarded as the limit of an infinite number of small parabolic segments. Formally, it can be shown, following Elmer's construction \cite{Elmer1980,winston:nonimaging}, that both the reflector and the light curve are connected by the proposed transformations. 
In fact, for continuous reflectors (or segments) that are purely converging or diverging, a polar diagram of the resulting radiant intensity can be generated from the light curve $(\bar{x},\bar{y})$. Let these coordinates be parametrized by a variable $\tau$. The length of a segment of this curve that spans a given infinitesimal range of directions is proportional to the local radius of curvature. This length, multiplied by an obliquity factor and divided by the distance to the origin, gives the infinitesimal angle subtended by this segment. The expression in polar coordinates is then given by
\begin{align}\label{eq:polar_diagram}
(I,\theta)&=\left[\frac{\dot{\bar{x}}^2+\dot{\bar{y}}^2}{\bar{x}^2+\bar{y}^2}\frac{|\bar{x}\dot{\bar{y}}-\dot{\bar{x}}\bar{y}|}{|\dot{\bar{x}}\ddot{\bar{y}}-\ddot{\bar{x}}\dot{\bar{y}}|},2\arctan{\left(\frac{\dot{\bar{y}}}{\dot{\bar{x}}}\right)}\right]\nonumber\\
&=\left[\frac{|\dot{\bar{z}}|^2{\rm Im}(\dot{\bar{z}}\bar{z}^*)}{|\bar{z}|^2{\rm Im}(\ddot{\bar{z}}\dot{\bar{z}}^*)},2\arg(\dot{\bar{z}})\right],
\end{align}
where the dot denotes a derivative in $\tau$. 
Note that $I$ diverges at the light curve's inflection points, which correspond to far-field caustics. These points typically separate segments that are convergent (which form caustics between the reflector and the far field) and divergent (which form no real caustics).

\begin{figure}[htpb]
\centering
\includegraphics[width=0.93\linewidth]{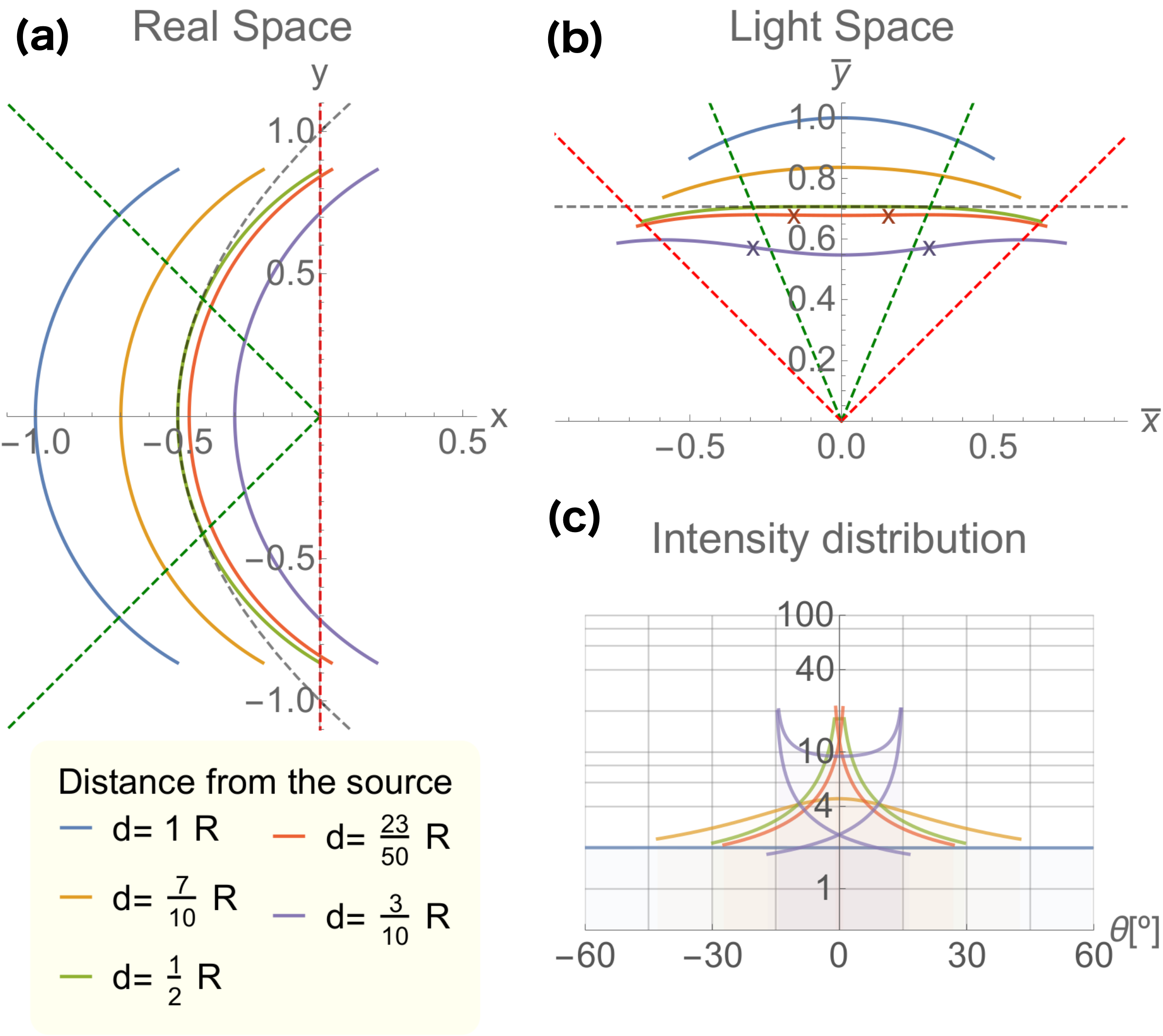}
\caption{\footnotesize (a) Circular reflector of radius $R$ at six distances $d$ from the source. 
(b) Their corresponding light curves. Within $\pm 45^\circ$ of the $x$-axis, (green dashed lines), 
this profile at  $d=R/2$ has good performance paraxially as a collimator, 
with a maximum deviation of $\pm 1.8^\circ$. Collimation degrades at higher angles, e.g. at $\pm 90^\circ$ (red dashed lines) the deviation is of $\pm 15^\circ$.
For a small deviation of $R/20$, the collimation range is extended to $\approx67.5^\circ$. For closer positions there is clear variation from divergent to convergent behavior, the crosses indicating the inflection points (far-field caustics). (c) 
Logarithmic radiant intensity diagrams. These become multivalued for non-purely convergent/divergent reflectors due to the appearance of far-field caustics (cusps on the plot).
(See {\bf Visualization 2} for more detail on this circular reflector as well as for the corresponding plots for reflectors with other conic sections.)
}
\label{fig:collimators}
\end{figure}

As an example of a continuous, smooth reflector, consider a circular 
reflector of radius $R$. 
This reflector is paraxially equivalent to an ideal parabolic collimator when the light source is placed half a radius away from its center ($f=R/2$). 
Figure~\ref{fig:collimators} shows (red curves) that the light curve in this case is considerably flat within the central region, mimicking the straight line for the ideal parabolic collimator (dashed line) within a range of source angles. 

This light curve is concave away from the central region, indicating convergent behavior. The convergence at the edges can be alleviated by slightly reducing the distance between the source and the reflector (essentially using defocus to balance spherical aberration) at the cost of introducing divergent behavior near the axis. 
This can be appreciated from the light curves in Fig.~\ref{fig:collimators}(b) for two locations closer than the collimating paraxial position. 
On the other hand, the behavior is purely convergent for circular reflectors further away from the paraxial collimating position, for which Eq. (\ref{eq:polar_diagram}) provides a polar diagram, as shown in Fig.~\ref{fig:collimators}(c). 
The corresponding behavior of other conic sections at different displacements is shown in {\bf Visualization 2}.

\begin{figure}[htbp]
\centering
\includegraphics[width=\linewidth]{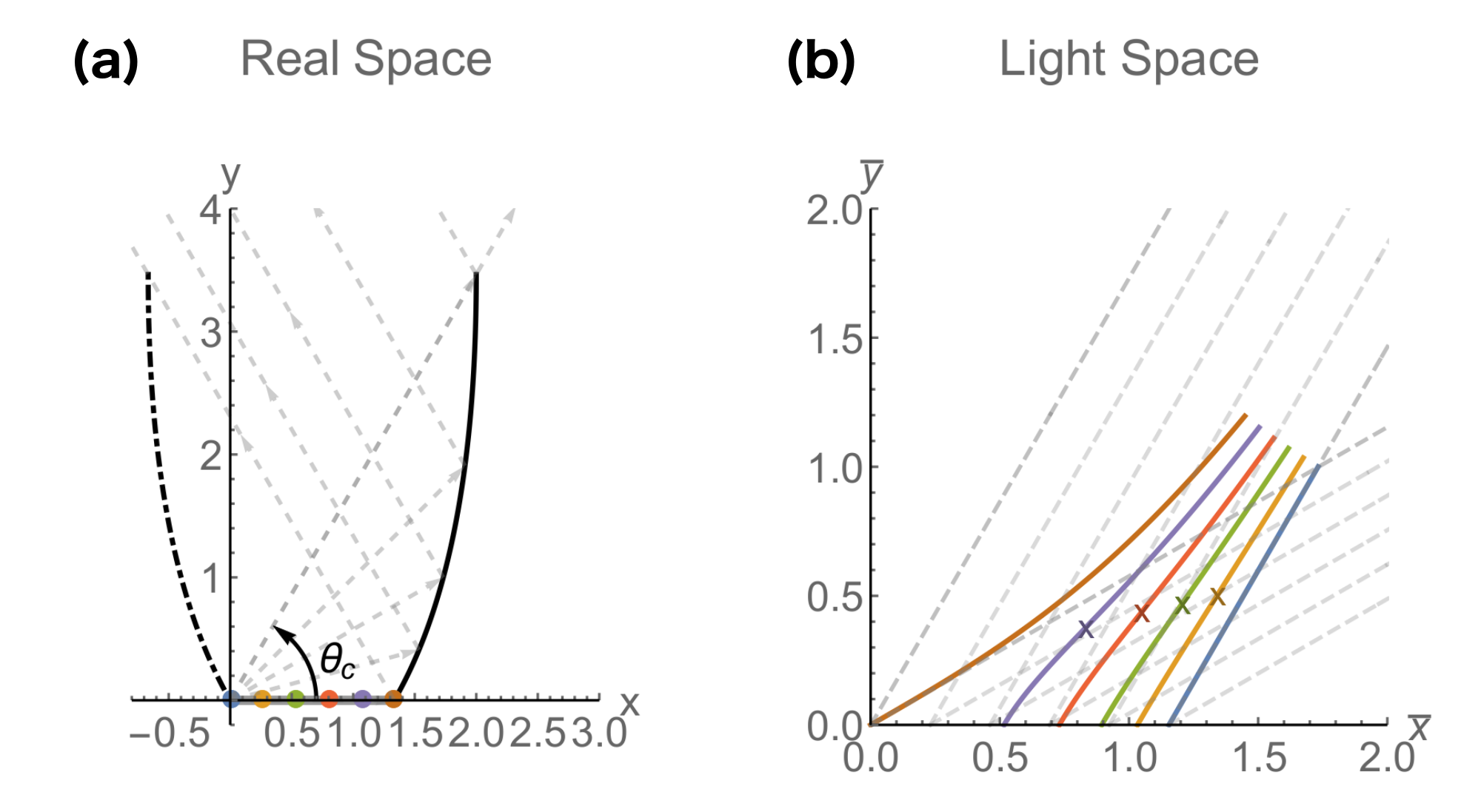}
\caption{(a) A CPC reflector ($\theta_c=\pi/3$, $f=1$) that collimates light from the edge source point along $\pi-\theta_c$, and six points along the extended source. (b) Light curves for these six source points. The gray-dashed lines 
indicate the directions for the acceptance angles $\theta_c$ and $\pi-\theta_c$. (See {\bf Visualization 3}.)
}
\label{fig:CPC_results}
\end{figure}

Let us finish with an example corresponding to an extended source/target: the 
Compound Parabolic Concentrator (CPC) \cite{Winston:70,WINSTON:74}, which achieves the ideal concentration limit \cite{Minano:84_cylindrical,Winston:18}. 
A CPC consists of a planar source/concentrator and two parabolic reflectors that redirect the light from/to the opposite end of the source/concentrator at $\pi\mp\theta_c$, where $\theta_c$ is the acceptance angle, i.e.,  the minimum angle from one of the corners of the source/concentrator to the upper end of the parabolic reflector (see Fig. \ref{fig:CPC_results}). 
By symmetry, 
only  one of the reflectors needs to be considered. The procedure is as follows: 
the origin is translated 
along the extended source/concentrator, and the map $\bar{z}=\sqrt{z}$ 
of the reflector 
reveals 
the direction in which light is directed after reflection. 
Figure~\ref{fig:CPC_results} shows the results for six points of the 
concentrator and an acceptance angle of $\theta_c=\pi/3$. The first point, corresponding to the opposite end of the concentrator, leads to a perfectly straight light line at $\theta_c/2$. As the point approaches the reflector, the direction angles change mostly for small source angles (small $\bar{y}$) to a divergent geometry, whereas for larger source angles (near the edge of the reflector) the light is directed at angles close to the acceptance angle. The direction for all the positions is bounded by $\theta_c/2$, which is achieved only at the initial position. This was to be expected, since the construction of the CPC can be derived using the edge-ray technique \cite{Winston:18}.

In summary, the maps $\sqrt{z}$ and $\bar{z}^2$ provide a new, simplified understanding of the relation between the reflector's shape and the light direction distribution after reflection, for both segmented and continuous reflectors. Light space can be used to design reflectors starting from their light curves, which are particularly simple for piecewise parabolic reflectors. 
Furthermore, from this new graphic representation, general reflector characteristics can be grasped easily even without explicitly tracing rays. For example, convergent/divergent/collimating geometries are mapped onto the convexity/concavity/straightness of the light curve. While the light curve conveys much of the same information as many rays would,  it does not fully replace them. For instance, one limitation of light curves is that they do not provide a direct visualization of cases when rays hit a reflector more than once (see {\bf Visualization 3}).

This approach is based on point-like sources (placed at the origin) which is a useful approximation in many cases. However, it can be applied also to extended sources by treating each point as a point source and translating the reflector before applying the map. This method is currently limited to systems whose symmetries reduce the problem to 2D, so that skew rays cannot yet be considered. Future work will focus on generalization to 3D. 
Another possible extension is to reflectors that generate desired irradiances at finite distances, for which the map should not only reveal the direction of light but also the distance at which it focuses.

\section*{Acknowledgements}
 We thank Mark Dennis, Jannick Rolland, Bill Cassarly and Cristina Canavesi for separate discussions whose intersection led to this idea.

\section*{Funding}
 University of Rochester's PumpPrimer II Award (OP212647),  CONACYT fellowship (LAC), Excellence Initiative of Aix-Marseille University-A*MIDEX, a French ``Investissements d'Avenir'' programme (MAA).


\end{document}